# Effects of Lanthanides on the Structure and Oxygen Permeability of Ti-doped Dual-phase Membranes

*Chao Zhang, Zaichen Xiang, Lingyong Zeng, Peifeng Yu, Kuan Li, Kangwang Wang, Longfu Li, Rui Chen, Huixia Luo\**

School of Materials Science and Engineering, State Key Laboratory of Optoelectronic Materials and Technologies, Guangdong Provincial Key Laboratory of Magnetoelectric Physics and Devices, Key Lab of Polymer Composite & Functional Materials, Sun Yat-Sen University, No. 135, Xingang Xi Road, Guangzhou, 510275, P. R. China.

*\*Corresponding author/authors complete details (Telephone; E-mail:) (+0086)-2039386124*

*luohx7@mail.sysu.edu.cn*

**ABSTRACT:**

The trade-off effect of the oxygen permeability and stability of oxygen transport membranes (OTMs) still exists in working atmospheres containing $CO_2$. Herein, we reported a new series of 60 wt%$Ce_{0.9}Ln_{0.1}O_{2-\delta}$-40wt%$Ln_{0.6}Sr_{0.4}Fe_{0.9}Ti_{0.1}O_{3-\delta}$ (CLnO-LnSFTO, Ln = La, Pr, Nd, Sm, Gd, Tb) dual-phase OTMs by selecting different Ln elements based on the reported highly stable Ti-doped CPrO-PrSFTO. The effects of different Ln elements on the structure and oxygen permeability of Ti-doped dual-phase OTMs were systematically studied. Basically, as the atomic number of Ln elements increases, the unit cell parameters of both the fluorite phase and the perovskite phase become smaller. The unit cell volume and spatial symmetry of the perovskite phase are reduced, resulting in a reduction in oxygen permeability. The optimal CLaO-LaSFTO showed $J_{O_2}$ of 0.60 and 0.54 mL min$^{-1}$ cm$^{-2}$ with He and $CO_2$ sweeping at 1000 ºC, respectively. In addition, all CLnO-LnSFTO OTMs could work for more than 100 hours with no significant performance degradation in a $CO_2$ atmosphere, maintaining excellent stability. This work explores candidate OTM materials for $CO_2$ capture and oxygen separation, as well as provides some ideas for addressing the trade-off effect.

**Keywords:** Oxygen transport membrane, Lanthanides, Titanium containing, Gas separation, $CO_2$ resistance

# 1. Introduction

Dual-phase oxygen transport membranes (OTM), including fluorite oxide and perovskite oxide, have been widely studied in oxygen separation, membrane reactors, solid oxide fuel cells (SOFC), etc [1-6]. However, there is still a trade-off effect in the oxygen permeability and stability of OTMs in working atmospheres containing $CO_2$ [7].

Alkaline earth metal elements (Ba, Sr, etc.) are often selected in the A site of the perovskite phase ($ABO_3$) because of their lower ionic valence and larger ionic radius. The larger unit cell volume and higher oxygen vacancy concentration allow materials containing these elements to exhibit higher oxygen permeability [2,8]. However, these ions also have lower relative acidity [9]. According to the Lewis acid-base theory, they are more likely to react with $CO_2$ to produce carbonate heterophases, thereby destroying the structure and oxygen conductivity of the material. $Ca^{2+}$ has been used to replace $Ba^{2+}$ or $Sr^{2+}$ because of the larger generation energy and lower decomposition temperature of its carbonate, showing better stability but poor performance [10,11]. Using high-valent ions ($Ti^{4+}$, $Nb^{5+}$, etc.) to replace Fe ions at the B site of $SrFeO_3$ partially can significantly improve stability [12,13]. A small amount of $Ti^{4+}$ doping has been reported to simultaneously improve oxygen permeability by generating more oxygen vacancies [14]. However, the fixed valence state prevents these ions from participating in the negligible polaron conduction of Fe-O-Fe, which is detrimental to the electronic conductivity and oxygen permeability of perovskite oxides. This makes it difficult to further improve performance by doping high-valent ions.

Replacing $Sr^{2+}$ with an appropriate amount of lanthanide (Ln) elements can also improve the ionic conductivity and stability of the perovskite phase [15,16]. Different cationic charges and radii have different effects on structure and performance [10,17-20]. For example, in the $Ln_{0.4}Sr_{0.6}Co_{0.8}Fe_{0.2}O_{3-\delta}$ (Ln = La, Pr, Nd, Sm, Gd) system, Nd-containing materials exhibited the highest conductivity, followed by La-containing and Pr-containing materials [18]. In addition, doping different Ln elements in the fluorite $CeO_2$ can also bring higher ionic conductivity and oxygen vacancy concentration, often used as the oxygen ion transport phase of dual-phase membranes [20]. In dual-phase $Ce_{0.9}Ln_{0.1}O_{2-\delta}$-$Ln_{0.6}Ca_{0.4}FeO_{3-\delta}$ (Ln = La, Pr, Nd, Sm), the membranes containing La or Pr performed better than that containing Nd or Sm [10]. This may be due to the latter's weaker sintering performance and the precipitation of heterogeneous phases after sintering, indicating that different Ln elements have an impact on the structure and performance of dual-phase OTMs.

In this work, we prepared a new series of 60 wt%$Ce_{0.9}Ln_{0.1}O_{2-\delta}$-40wt%$Ln_{0.6}Sr_{0.4}Fe_{0.9}Ti_{0.1}O_{3-\delta}$ (CLnO-LnSFTO, Ln = La, Pr, Nd, Sm, Gd, Tb) dual-phase OTMs by selecting different Ln elements based on the reported $Ce_{0.9}Pr_{0.1}O_{2-\delta}$-$Pr_{0.6}Sr_{0.4}Fe_{0.9}Ti_{0.1}O_{3-\delta}$, which exhibited good stability and oxygen permeability. A small amount of Ti doping could improve the stability of the material in atmospheres containing $CO_2$, and the suitable doping of Ln elements could further improve the oxygen ionic conductivity and oxygen permeability. This work systematically studied the effects of different Ln elements on the structure and oxygen permeability of highly stable dual-phase Ti-doped membranes without the precipitation of impurities similar to those in Ca-containing membranes.

## 2. Experimental
### 2.1. Synthesis and characterization

The modified Pechini method was used to synthesize CLnO-LnSFTO dual-phase powders [12,21]. The crystal structure of powders was examined using room temperature X-ray diffraction (XRD, Rigaku MiniFlex 600, Cu Kα) within the 2θ range of 10 - 100°. Crystal structure diagrams were generated using the Vesta software based on the Rietveld refinements [22]. The powders were then pressed into disks using a stainless-steel mold at ~ 150 MPa. Subsequently, the membranes were sintered at 1450 °C in air for 5 hours, with a ramp of 1 °C min$^{-1}$. The surface morphology and element composition of membranes were characterized by scanning electron microscopy (SEM, ZEISS EVO MA10), backscattered scanning electron microscopy (BSEM), and energy dispersive X-ray spectroscopy (EDXS).

## 2.2. Calculation of the tilting angle

Due to the complicated mixed chemical valence in the perovskite phase, such as Pr (+3/+4), Co (+2/+3), and Fe (+3/+4), it is difficult to calculate the tolerant factor via the ionic radius. Thus, a new parameter, tilting angle $\Phi$, was introduced in the perovskite, describing the twist distortion of the B–O octahedral, with the ideal one being along the three-fold axis [23]. When $\Phi$ is smaller, the octahedral shape tends to be the ideal one; alternatively, it tends to get converted to a more stable cubic structure. The tilting angle $\Phi$ is calculated according to **Eq. (1)**:

$$\Phi = \cos^{-1}\left(\frac{\sqrt{2}a^2}{bc}\right) \tag{1}$$

where $a$, $b$, and $c$ denote the small, medium, and large parameters of the unit cell, respectively. This equation is only valid for perovskites with *Pbnm* space groups.

## 2.3. Oxygen permeation evaluation

All oxygen permeation tests were conducted on a reported self-assembled high-temperature device [24]. OTMs were sealed on top of the corundum tube with the industrial high-temperature adhesive (Huitian 2767) after being polished to a thickness of 0.6 mm. The effective working area, calculated with the inner diameter of corundum tubes and confirmed by measuring the spent membranes, is approximately 0.65 cm$^2$. The feed gas consisted of synthetic air (150 mL min$^{-1}$, 21% O$_2$, and 79% N$_2$). On the other side of the membranes, a mixture of gas (1 mL min$^{-1}$ Ne + 49 mL min$^{-1}$ He/CO$_2$) was used for sweeping. The flow rates mentioned above were determined using mass flow control meters. The components of exhaust gases from the sweep side were subjected to gas chromatography analysis (GC, PANNA-A60, China). Samples with an oxygen leakage ratio lower than 5 % were selected for testing, and the oxygen permeation flux $J_{O_2}$ (mL min$^{-1}$ cm$^{-2}$) could be calculated by deducting the leakage (**Eq. (2)**) [25]:

$$J_{O_2} = \frac{F}{S} \times \left(C_{O_2} - \frac{C_{N_2}}{4.02}\right) \tag{2}$$

Where $F$ represents the total gas flow rate into the GC, and S refers to the effective working area of OTMs. $C_{N_2}$ and $C_{O_2}$ represent the concentrations of N$_2$ and O$_2$

## 3. Results and discussion
### 3.1. Phase structure of powders and membranes

**Fig. 1** shows the XRD refinements with the Rietveld model for the as-obtained CLnO-LnSFTO (Ln = La, Pr, Nd, Sm, Gd, Tb) composite powders. The data reveal that all our investigated dual-phase powders, after calcined in 950 ºC for 12 hours, only consist of the cubic fluorite phase (space group: $Fm\bar{3}m$, No. 225) and the pseudo-cubic ($a \times b \times c \approx \sqrt{2}\,a_c \times \sqrt{2}\,a_c \times 2a_c$, where $a$, $b$, $c$ represent unit cell parameters of the orthorhombic structure, and $a_c$ represents the unit cell parameter of cubic structure [26]) perovskite phase (space group $Pbnm$, No. 62). **Fig. S1** illustrates the structure of CLaO and LaSFTO, generated with VESTA [22]. The unit cell parameters for our investigated compounds are listed in **Table 1**. For comparison, the converted cell parameters and volumes of powders are further shown in **Fig. 2a** and **2b**. The parameters and volumes generally decreased with the increases in the atomic numbers, which can be attributed to the decrease of the corresponding ionic radii due to the lanthanide shrinkage effect (see **Table S1**, extracted from the Jia's table [27]). In particular, the cell parameter value for the fluorite phase in the CPrO-PrSFTO is smaller than that in CNdO-NdSFTO and CSmO-SmSFTO. It can be explained that Pr and Tb ions are multivalent, such as +3 and +4. $Pr^{4+}$ (96 pm) and $Tb^{4+}$ (88 pm) have much smaller ionic radii than $Pr^{3+}$ (112.6 pm) and $Tb^{3+}$ (104 pm) in the eight coordinations, whereas other rare-earth elements (La, Nd, Sm, Gd) only exhibit a fixed valence (+3) in the eight coordinations. This change trend is in agreement with that in the single-phase fluorite phase $Ce_{1-x}Ln_xO_{2-\delta}$ system (Ln = Gd, La, Tb, Pr, Eu, Er, Yb, Nd) reported by Balaguer et al. [20].

In order to observe the evolution of the cell parameters for perovskite phases intuitively, we divide $a$, $b$ by $\sqrt{2}$ and $c$ by 2 (Based on the pseudo-cubic structure, $a \times b \times c \approx \sqrt{2}\,a_c \times \sqrt{2}\,a_c \times 2a_c$). As shown in **Fig. 2a**, the difference in converted cell parameters ($a/\sqrt{2}$, $b/\sqrt{2}$, $c/2$) of perovskite phases increases with the increase of atomic numbers. Specifically, the converted cell parameters are almost the same in the case of CLaO-LaSFTO, whereas they are very different in the CTbO-TbSFTO compound. That is to say, the symmetries of the perovskite phase in obtained powders are getting worse with the increases in atomic numbers of rare earth elements. Also, the tilting angle $\Phi$ is an important geometrical parameter to evaluate the twist distortion of the $BO_6$ octahedral in the orthorhombic perovskite structure [28]. As the $\Phi$ value gets closer to 0, the distorted $BO_6$ octahedron will be closer to the ideal normal octahedron. In other words, the crystal structure is more likely to change to a stable cubic structure in high temperatures for easier oxygen permeation. The calculated $\Phi$ values for our investigated compounds are summarized in **Table 1,** and they gradually increase from 3.66° to 13.97° with the atomic number increasing, indicating that the crystal symmetry becomes worse.

XRD was used to further characterize the crystal structure of the membranes after sintering at 1450 °C for 5 hours. The refined results using the retiveld model (**Fig. S2**) indicate that all diffraction peaks belong to the fluorite phase and the perovskite phase. Moreover, the mass ratio of the two phases was close to the theoretical value of 60 wt.%. No third-phase impurities or element segregation appeared in all membranes studied, which is consistent with the reported $Ce_{0.9}Pr_{0.1}O_{2-\delta}$-$Pr_{0.6}Sr_{0.4}Fe_{0.9}Ti_{0.1}O_{3-\delta}$. In particular, there is no such problem that the perovskite phase partially decomposes into the $SrPr_4Ti_5O_{17}$ after sintering when Fe is completely substituted with Ti, indicating that obtained dual-phase OTMs containing different Ln elements have good stability during the sintering process [14].

**Table 2** shows the unit cell parameters of the sintered dual-phase membranes derived from XRD refinements. The influence of Ln elements on the values of unit cell parameters of the two phases and the symmetry of the perovskite phase is basically consistent with the results for powders. Compared with the powder samples, the XRD results of all membranes show that the tilting angle ($\Phi$) of the perovskite phase is larger, indicating that the $BO_6$ octahedral structure becomes more distorted. In other words, the spatial symmetry of the perovskite phase becomes worse after sintering. This can be due to two factors. First, during the sintering process of dual-phase MIEC OTMs, the perovskite phase with a larger thermal expansion coefficient would undergo internal compressive strain [29]. This strain leads to more distorted $BO_6$ and worse symmetry of the perovskite phase. Another possible factor is that the formation of oxygen vacancies is not uniform along the three axes during the sintering process of the orthorhombic perovskite phase. For example, oxygen vacancies tend to form at one corner of the *ab* plane in Fe-doped $SrCo_{0.95}Ti_{0.05}O_{3-\delta}$ [30]. The unit cell parameters of the *b* and *c* axes increase more, so $\Phi$ increases as well after sintering, resulting in a decrease in spatial symmetry. **Fig. 3** shows the close-up XRD patterns and annotated diffraction peaks near 32°. It is worth noting that whether in powders or membranes, the relative diffraction intensity of the perovskite phase decreases significantly compared with that of the fluorite phase as the atomic number of the Ln element increases. This is different from the enhanced considerably diffraction intensity of the perovskite phase in Pr-based dual-phase membranes with high Ti content after sintering [14], indicating that the poor crystallinity in the powders does not cause it due to the low calcination temperature. The decrease in symmetry of the unit cell of the perovskite phase can explain this.

Moreover, for pseudo-cubic perovskite (space group *Pbnm*, No. 62), the diffraction peak positions of (200), (020), and (112) crystal planes in the ideal cubic structure should coincide at about 32°. In other words, the closer the diffraction peaks of the three crystal planes are, the closer the crystal structure is to the ideal cubic structure and the higher the symmetry is. In the XRD patterns of membranes, the three diffraction peaks of the perovskite phase in CLaO-LaSFTO, CPrO-PrSFTO, and CNdO-NdSFTO are closer, and they exhibit higher symmetry. In addition, the diffraction peaks of the membranes after sintering are significantly more separated, especially the CGdO-GdSFTO and CTbO-TbSFTO membrane shows significant peak-splitting. This further shows that the symmetry of the perovskite phase in the sample decreases after sintering, which is consistent with the analysis results of $\Phi$. Furthermore, the unit cell parameters and volumes of dual-phase membranes generally decrease with the increase of the atomic number of Ln elements, except for CPrO-PrSFTO (**Fig. 2c and 2d**), which is consistent with the results of the powder samples.

**3.2. Membrane surface morphology**

The surface morphology and phase structure of sintered membranes were characterized by SEM, BSEM, and EDXS. There was no significant difference in the surface morphology of membranes containing different Ln elements (**Fig. 4**), and only a small number of blind holes were observed in all membranes without large through holes or cracks. In the SEM images (**Fig. 4a-4f**), the grains are densely arranged, and the grain boundaries are clearly visible, which indicates that the grains of the two phases grow well and no solid-phase reaction occurred. In the BSEM images (**Fig. 4g-4l**), there are two areas of different brightness, indicating that the membranes are mainly composed of CLnO and LnSFTO phases. The brightness contrast between areas belonging to different phases represents the difference in average atomic mass, where the brighter area and the darker area represent fluorite CLnO and perovskite LnSFTO, respectively. The uniform distribution is beneficial to the exchange of oxygen ions and electrons between the two phases and enhances the oxygen permeation in the dual-phase OTMs [31]. The grain sizes of the two phases in CLnO-LnSFTO are similar, indicating that the grain growth states of all compounds are similar at the selected sintering temperature.

Furthermore, although Ti-containing OTMs require a higher sintering temperature than materials containing sintering aids (CaO, CoO, CuO, etc.), there is no decomposition of the perovskite phase in membranes sintered at the required temperature. The oxides of the doping elements did not precipitate, and the membranes had good compactness. Partial doping of Ti to replace Fe also reduces the thermal expansion coefficient of the perovskite phase [32], and there are no microcracks that are easy to appear on the surface of Co-containing OTMs [33].

In order to further observe the phase distribution on the membrane surfaces, EDXS

was used to characterize the distribution of elements (**Fig. 5**). The unique elements of the two phases (Ce in the fluorite phase and Fe/Sr in the perovskite phase) in all compounds showed obvious complementary distributions. The theoretical values of each element content were calculated based on the mass fraction of the two phases and compared with the experimental values (**Table S2**), showing good consistency. The above results show that we successfully prepared the designed dual-phase OTMs.

**3.3. Oxygen permeation test**

To explore the oxygen permeability of CLnO-LnSFTO dual-phase OTMs, we focused on oxygen permeation tests at different temperatures. All sintered membranes were ground and polished to the same thickness of 0.6 mm before testing. **Fig. 6a** and **6b** show curves of oxygen permeation fluxes ($J_{O_2}$) versus temperature through CLnO-LnSFTO dual-phase OTMs under pure He and pure $CO_2$ sweeping, respectively. As the temperature increased, the bulk diffusion and surface exchange of oxygen in OTMs were significantly enhanced, resulting in improved oxygen permeability. The optimal CLaO-LaSFTO showed $J_{O_2}$ of 0.60 and 0.54 mL min$^{-1}$ cm$^{-2}$ with He and $CO_2$ sweeping at 1000 °C, respectively. As the permeation process is controlled by a mixture of surface exchange and bulk diffusion, there may be a conversion between these two situations at 800-900 °C for our 0.6-mm-thickness membranes [34]. In addition, the possible phase transition of the perovskite phase [35] and the small leakage of membranes cause changes in the activation energy ($E_a$) around 900 °C. Due to the focus on comparing the oxygen permeability at high temperatures, we mainly selected data above 900 °C when calculating the activation energy. **Fig. S3a** and **S3b** are Arrhenius plots based on $J_{O_2}$ in He/$CO_2$ at 900-1000 °C, respectively. The apparent activation energy obtained by fitting further shows that the oxygen permeation of dual-phase membranes has obvious thermal activation characteristics and is easier to proceed at high temperatures. Most membranes exhibit higher $E_a$ with $CO_2$ sweeping, which is consistent with stronger surface adsorption of $CO_2$. The exception exhibited by Sm-doped OTMs is due to a slight decrease in sealing performance after switching the sweep gas into $CO_2$ and permeability test during the cooling process. It resulted in a higher oxygen partial pressure on the sweep side and reduced the oxygen partial pressure gradient in a $CO_2$ atmosphere. Arrhenius plots (**Fig. S3b**) showed the oxygen permeability increased slowly with temperature rise, exhibiting abnormally low $E_a$.

Comparing the oxygen permeability of CLnO-LnSFTO OTMs at 1000 °C (**Fig. 6c**), La-containing dual-phase membranes showed the highest permeability in both atmospheres, followed by those containing Pr and Nd. Overall, the oxygen permeability of the CLnO-LnSFTO weakens as the atomic number of the Ln element increases, which is consistent with the changing trend of the activation energy ($E_a$). This can be explained from the structural aspect. As the atomic number increases, the ionic radius of $Ln^{3+}$ decreases due to the lanthanide shrinkage effect. The fitting results in **Table 2** show that the unit cell parameters also gradually decrease except for CPrO-PrSFTO, which reduces the free volume within the unit cell and the possibility of generating oxygen vacancies [7,15]. In addition, the spatial symmetry of the crystal will also affect the oxygen permeability. It is generally believed that the cubic phase with high symmetry is more conducive to oxygen transport because oxygen ions can be transported three-dimensionally in the cubic phase but are directionally limited in the orthorhombic phase. For example, oxygen ions can easily migrate on the *ab* plane while difficulty along the *c* axis in orthorhombic $La_{0.64}(Ti_{0.92}Nb_{0.08})O_{2.99}$ compound [36]. Therefore, in our XRD refinements, the doping of Ln elements with larger atomic numbers (Gd, Tb, etc.) weakens the symmetry of the orthorhombic perovskite phase and also leads to a reduction in oxygen permeability.

It is worth noting that the unit cell parameters of CPrO-PrSFTO are smaller than those of CNdO-NdSFTO, but their oxygen permeability is similar. This may be attributed to the non-negligible electronic conductivity of CPrO, which is beneficial to the charge exchange between the two phases. Due to the mixed valence states of the Pr element ($Pr^{3+}$, $Pr^{4+}$), some small polarons in fluorite CPrO can jump between different valence states, thereby increasing the electronic conductivity [37,38]. For example, under an oxygen partial pressure gradient of 1/0.21 bar, the electronic conductivity of Ce$_{0.8}$Pr$_{0.2}$O$_{2-\delta}$ is 0.021 S cm$^{-1}$, which is the same order of magnitude as the ionic conductivity (0.077 S cm$^{-1}$) [39]. Meanwhile, the electronic conductivity of most other CLnO oxides is negligible due to the fixed valence state (+3) of the lanthanides. For example, Ce$_{0.9}$Gd$_{0.1}$O$_{2-\delta}$ exhibited the electronic conductivity of 10$^{-3.8}$ S cm$^{-1}$ at 800 °C [40]. Since the overall electronic conductivity of dual-phase OTMs is much higher than the ionic conductivity, the Wagner equation (**Eq. (3)**) could be simplified to **Eq. (4)** and then transformed into **Eq. (5)** [41]. The ionic conductivity of the CLnO-LnSFTO dual-phase membranes can be approximately calculated by **Eq. (5)**. Consistent with the oxygen permeability, the La-containing membranes showed the optimal ionic conductivity (**Fig. 6d**). Pr-containing membranes were slightly better than those containing Nd, supporting our previous discussion.

$$J_{O_2} = \frac{RT}{16F^2L} \int_{P_l}^{P_h} \frac{\sigma_{ion}\sigma_e}{\sigma_{ion}+\sigma_e} \mathrm{d}(\ln P_{O_2}) \tag{3}$$

$$J_{O_2} = \frac{RT}{16F^2L} \sigma_{ion} \ln \frac{P_h}{P_l} \tag{4}$$

$$\sigma_{ion} = J_{O_2} \frac{16F^2L}{RT(\ln P_h - \ln P_l)} \tag{5}$$

Where $R$, $T$, $F$, and $L$ represent the gas constant, temperature, Faraday constant, and membrane thickness, respectively. $P_{O_2}$ represents the partial pressure of oxygen. $P_l$ and $P_h$ are the lower and higher $P_{O_2}$ of two sides of the membranes, respectively. $\sigma_{ion}$ is the ionic conductivity and $\sigma_e$ is the electronic conductivity.

In addition to the influence depending on atomic number, Ln elements can also affect the sintering performance of dual-phase membranes. For Ca-containing Ce$_{0.9}$Ln$_{0.1}$O$_{2-\delta}$-Ln$_{0.6}$Ca$_{0.4}$FeO$_{3-\delta}$ (Ln = La, Pr, Nd, Sm), Nd-doped and Sm-doped Ln$_{0.6}$Ca$_{0.4}$FeO$_{3-\delta}$ would decompose into different proportions of perovskite phases. This hinders oxygen migration, resulting in a decrease in oxygen permeability [10]. There is no severe such phenomenon for our Ti-containing OTMs. However, the fluorite phase occupies the surface of the Gd-doped and Tb-doped OTMs after sintering, and the perovskite phase is more dispersedly distributed instead of forming a continuous conductive network. This would influence the electronic conductivity and, thus, the oxygen permeation flux.

**3.4. Stability of materials**

To initially study the structural stability of the dual-phase material, we heat-treated the powder samples in pure Ar and pure $CO_2$ for 24 hours before sintering. XRD results showed that the powders in both atmospheres did not undergo phase transformation after heat treatment at 800 ºC, 900 ºC, and 1000 ºC (**Fig. S4**). All diffraction peaks could be attributed to the cubic fluorite phase and orthorhombic perovskite phase. There were no other impurity peaks, and this was the same as fresh powders. It is worth noting that no carbonate formation was observed in all powders in a $CO_2$ atmosphere. Some OTM materials containing Sr and Ca are reported to generate carbonate in $CO_2$, resulting in oxygen permeability reduction and structural damage [42,43]. In contrast, our powders remained stable in pure $CO_2$, demonstrating that doping with different Ln elements did not damage the excellent $CO_2$ resistance of Ti-doped dual-phase OTMs.

Further, the performance stability in the long-term oxygen permeation test is demonstrated in **Fig. 7**. All CLnO-LnSFTO OTMs worked stably for 50 and 100 hours, respectively, under He and $CO_2$ sweeping at 1000 ºC, with basically no performance degradation. For some materials like CeGdO-BaSrCoFeO [2], when the sweep gas was changed from He to $CO_2$, the oxygen permeation flux quickly decreased to close to 0. However, our CLnO-LnSFTO OTMs only showed a small performance drop after switching. This could be attributed to the partial occupation of surface active sites by $CO_2$ rather than the destruction of the oxygen transport pathway caused by carbonate [7,44]. **Fig. S5** and **S6** show the SEM-EDXS images of the feed side and sweep side of spent membranes, respectively after long-term oxygen permeation tests. There were no third-phase impurity grains on the surface of all CLnO-LnSFTO membranes, and there was no segregation of some aspects in element mapping. This showed that our dual-phase OTMs maintained a uniform distribution of the two phases and a continuous path for charge exchange in tests, which was beneficial to the oxygen permeation process [45]. The element content on both sides characterized by EDXS (**Table S3**) was basically consistent with that of the fresh membrane, indicating that there was no large-scale transfer of cations between the surfaces and interior of materials. This was consistent with the reported CPrO-PrSFTO, indicating the excellent performance stability of CLnO-LnSFTO in high-temperature $CO_2$-containing working atmospheres.

## 4. Conclusions

A series of 60 wt%$Ce_{0.9}Ln_{0.1}O_{2-\delta}$-40wt%$Ln_{0.6}Sr_{0.4}Fe_{0.9}Ti_{0.1}O_{3-\delta}$ (CLnO-LnSFTO, Ln = La, Pr, Nd, Sm, Gd, Tb) dual-phase OTMs materials have been prepared successfully by the modified Pechini method. Different Ln elements affect the structure, oxygen permeability, and stability of Ti-doped dual-phase OTMs. Basically, as the atomic number of Ln elements increases, the unit cell parameters of both the fluorite phase and the perovskite phase become smaller. The unit cell volume and spatial symmetry of the perovskite phase are reduced, resulting in a decrease in oxygen ionic conductivity and oxygen permeability. The optimal CLaO-LaSFTO showed $J_{O_2}$ of 0.60 and 0.54 mL min$^{-1}$ cm$^{-2}$ with He and $CO_2$ sweeping at 1000 ºC, respectively. In addition, all CLnO-LnSFTO OTMs could work for more than 100 hours with no significant performance degradation in a $CO_2$ atmosphere, maintaining excellent stability. This indicates that CLnO-LnSFTO OTMs are potentially applicable materials for $CO_2$ capture and oxygen separation.

**Acknowledgment**

This work is supported by the Natural Science Foundation of China (No. 11922415, 12274471), Guangdong Basic and Applied Basic Research Foundation (No. 2022A1515011168), Guangzhou Science and Technology Programme (No. 2024A04J6415) and the State Key Laboratory of Optoelectronic Materials and Technologies (Sun Yat-Sen University, No. OEMT-2024-ZRC-02) . The experiments reported were conducted at the Guangdong Provincial Key Laboratory of Magnetoelectric Physics and Devices, No. 2022B1212010008. Lingyong Zeng was thankful for the Postdoctoral Fellowship Program of CPSF (GZC20233299) and

**Reference**

[1] T. Wang, Z. Liu, X. Xu, J. Zhu, G. Zhang, W. Jin, Insights into the design of nineteen-channel perovskite hollow fiber membrane and its oxygen transport behaviour, Journal of Membrane Science 595 (2020) 117600. https://doi.org/10.1016/j.memsci.2019.117600.

[2] J. Xue, Q. Liao, Y.Y. Wei, Z. Li, H.H. Wang, A $CO_2$-tolerance oxygen permeable $60Ce_{0.9}Gd_{0.1}O_{2-\delta}$–$40Ba_{0.5}Sr_{0.5}Co_{0.8}Fe_{0.2}O_{3-\delta}$ dual phase membrane, Journal of Membrane Science 443 (2013) 124-130. https://doi.org/10.1016/j.memsci.2013.04.067.

[3] H. Luo, H. Jiang, T. Klande, Z. Cao, F. Liang, H. Wang, J. Caro, Novel Cobalt-Free, Noble Metal-Free Oxygen-Permeable $40Pr_{0.6}Sr_{0.4}FeO_{3-\delta}$–$60Ce_{0.9}Pr_{0.1}O_{2-\delta}$ Dual-Phase Membrane, Chemistry of Materials 24 (2012) 2148-2154. https://doi.org/10.1021/cm300710p.

[4] L.J. Jia, G.H. He, Y. Zhang, J. Caro, H.Q. Jiang, Hydrogen Purification through a Highly Stable Dual-Phase Oxygen-Permeable Membrane, Angew. Chem. Int. Ed. 60 (2021) 5204-5208. https://doi.org/10.1002/anie.202010184.

[5] Z.P. Shao, S.M. Haile, A high-performance cathode for the next generation of solid-oxide fuel cells, Nature 431 (2004) 170-173. https://doi.org/10.1038/nature02863.

[6] Y.H. Liu, K. Kang, Z.F. Pan, C. Wang, K.T. Jiang, Y. Wang, Enhancing the oxygen reduction reaction activity of SOFC cathode via construct a cubic fluorite/perovskite heterostructure, Applied Surface Science 642 (2024) 158405. https://doi.org/10.1016/j.apsusc.2023.158405.

[7] C. Zhang, J. Sunarso, S. Liu, Designing $CO_2$-resistant oxygen-selective mixed ionic–electronic conducting membranes: guidelines, recent advances, and forward directions, Chem. Soc. Rev. 46 (2017) 2941-3005. https://doi.org/10.1039/C6CS00841K.

[8] Y. Zhu, D.D. Liu, H.J. Jing, F. Zhang, X.B. Zhang, S.Q. Hu, L.M. Zhang, J.Y. Wang, L.X. Zhang, W.H. Zhang, B.J. Pang, P. Zhang, F.T. Fan, J.P. Xiao, W. Liu, X.F. Zhu, W.S. Yang, Oxygen activation on Ba-containing perovskite materials, Sci. Adv. 8 (2022) eabn4072. https://doi.org/10.1126/sciadv.abn4072.

[9] N.C. Jeong, J.S. Lee, E.L. Tae, Y.J. Lee, K.B. Yoon, Acidity Scale for Metal Oxides and Sanderson's Electronegativities of Lanthanide Elements, Angew. Chem. Int. Ed. 47 (2008) 10128-10132. https://doi.org/10.1002/anie.200803837.

[10] S. Wang, L. Shi, M. Boubeche, X.P. Wang, L.Y. Zeng, H.Q. Wang, Z.A. Xie, W. Tan, H.X. Luo, Influence of Ln elements (Ln = La, Pr, Nd, Sm) on the structure and oxygen permeability of Ca-containing dual-phase membranes, Separation and Purification Technology 251 (2020) 117361. https://doi.org/10.1016/j.seppur.2020.117361.

[11] K. Efimov, T. Klande, N. Juditzki, A. Feldhoff, Ca-containing $CO_2$-tolerant perovskite materials for oxygen separation, Journal of Membrane Science 389 (2012) 205-215. https://doi.org/10.1016/j.memsci.2011.10.030.

[12] M.K. Liu, Z.W. Cao, W.Y. Liang, Y. Zhang, H.Q. Jiang, Membrane Catalysis: $N_2O$ Decomposition over $La_{0.2}Sr_{0.8}Ti_{0.2}Fe_{0.8}O_{3-\delta}$ Membrane with Oxygen Permeability, Chemie Ingenieur Technik 94 (2022) 70-77. https://doi.org/10.1002/cite.202100122.

[13] Z.G. Wang, N. Dewangan, S. Das, M.H. Wai, S. Kawi, High oxygen permeable and $CO_2$-tolerant $SrCo_xFe_{0.9-x}Nb_{0.1}O_{3-\delta}$ ($x$ = 0.1–0.8) perovskite membranes: Behavior and mechanism, Separation and Purification Technology 201 (2018) 30-40. https://doi.org/10.1016/j.seppur.2018.02.046.

[14] C. Zhang, L.Y. Zeng, P.F. Yu, K. Li, K.W. Wang, L.F. Li, Z.C. Xiang, H.X. Luo, Highly stable dual-phase $Ce_{0.9}Pr_{0.1}O_{2-\delta}$-$Pr_{0.6}Sr_{0.4}Fe_{1-x}Ti_xO_{3-\delta}$ oxygen transport membranes, Journal of Membrane Science 693 (2024) 122359. https://doi.org/10.1016/j.memsci.2023.122359.

[15] J. Lu, Y.M. Yin, J.W. Yin, J.C. Li, J. Zhao, Z.F. Ma, Role of Cu and Sr in Improving the Electrochemical Performance of Cobalt-Free $Pr_{1-x}Sr_xFe_{1-y}Cu_yO_{3-\delta}$ Cathode for Intermediate Temperature Solid Oxide Fuel Cells, Journal of The Electrochemical Society 163 (2016) F44. https://doi.org/10.1149/2.0181602jes.

[16] J.W. Yin, Y.M. Yin, J. Lu, C.M. Zhang, N.Q. Minh, Z.F. Ma, Structure and Properties of Novel Cobalt-Free Oxides $Nd_xSr_{1-x}Fe_{0.8}Cu_{0.2}O_{3-\delta}$ ($0.3 \leq x \leq 0.7$) as Cathodes of Intermediate Temperature Solid Oxide Fuel Cells, The Journal of Physical Chemistry C 118 (2014) 13357-13368. https://doi.org/10.1021/jp500371w.

[17] Y. Ren, R. Küngas, R.J. Gorte, C. Deng, The effect of A-site cation (Ln = La, Pr, Sm) on the crystal structure, conductivity and oxygen reduction properties of Sr-doped ferrite perovskites, Solid State Ionics 212 (2012) 47-54. https://doi.org/10.1016/j.ssi.2012.02.028.

[18] H.Y. Tu, Y. Takeda, N. Imanishi, O. Yamamoto, $Ln_{0.4}Sr_{0.6}Co_{0.8}Fe_{0.2}O_{3-\delta}$ (Ln = La, Pr, Nd, Sm, Gd) for the electrode in solid oxide fuel cells, Solid State Ionics 117 (1999) 277-281. https://doi.org/10.1016/S0167-2738(98)00428-7.

[19] Y.H. Chen, Y.J. Wei, H.H. Zhong, J.F. Gao, X.Q. Liu, G.Y. Meng, Synthesis and electrical properties of $Ln_{0.6}Ca_{0.4}FeO_{3-\delta}$ (Ln = Pr, Nd, Sm) as cathode materials for IT-SOFC, Ceramics International 33 (2007) 1237-1241. https://doi.org/10.1016/j.ceramint.2006.03.035.

[20] M. Balaguer, C. Solís, J.M. Serra, Structural–Transport Properties Relationships on $Ce_{1-x}Ln_xO_{2-\delta}$ System (Ln = Gd, La, Tb, Pr, Eu, Er, Yb, Nd) and Effect of Cobalt Addition, The Journal of Physical Chemistry C 116 (2012) 7975-7982. https://doi.org/10.1021/jp211594d.

[21] Y.H. Huang, C. Zhang, L.Y. Zeng, Y.Y. He, P.F. Yu, K. Li, H.X. Luo, Dual-phase Ga-containing $Ce_{0.9}Pr_{0.1}O_{2-\delta}$-$Pr_{0.6}Sr_{0.4}Fe_{1-x}Ga_xO_{3-\delta}$ oxygen transport membranes with high $CO_2$ resistance, Journal of Membrane Science 668 (2023) 121260. https://doi.org/10.1016/j.memsci.2022.121260.

[22] K. Momma, F. Izumi, VESTA 3 for three-dimensional visualization of crystal,

volumetric and morphology data, J. Appl. Crystallog. 44 (2011) 1272-1276. https://doi.org/10.1107/S0021889811038970.

[23] Y. Zhao, D.J. Weidner, J.B. Parise, D.E. Cox, Thermal expansion and structural distortion of perovskite — data for NaMgF3 perovskite. Part I, Physics of the Earth and Planetary Interiors 76 (1993) 1-16. https://doi.org/10.1016/0031-9201(93)90051-A.

[24] H.X. Luo, H.Q. Jiang, K. Efimov, F.Y. Liang, H.H. Wang, J. Caro, $CO_2$-Tolerant Oxygen-Permeable $Fe_2O_3$-$Ce_{0.9}Gd_{0.1}O_{2-\delta}$ Dual Phase Membranes, Industrial & Engineering Chemistry Research 50 (2011) 13508-13517. https://doi.org/10.1021/ie200517t.

[25] D.C. Li, X.P. Wang, W. Tan, Y.H. Huang, L.Y. Zeng, Y.Y. He, P.F. Yu, H.X. Luo, Influences of Al substitution on the oxygen permeability through 60 wt%$Ce_{0.9}La_{0.1}O_{2-\delta}$-40 wt%$La_{0.6}Sr_{0.4}Co_{1-x}Al_xO_{3-\delta}$ composite membranes, Separation and Purification Technology 274 (2021) 119042. https://doi.org/10.1016/j.seppur.2021.119042.

[26] R.D. Shannon, Revised effective ionic radii and systematic studies of interatomic distances in halides and chalcogenides, Acta Crystallographica Section A 32 (1976) 751-767. https://doi.org/10.1107/S0567739476001551.

[27] Y.Q. Jia, Crystal radii and effective ionic radii of the rare earth ions, Journal of Solid State Chemistry 95 (1991) 184-187. https://doi.org/10.1016/0022-4596(91)90388-X.

[28] S. Wang, L. Shi, M. Boubeche, H.Q. Wang, Z.A. Xie, W. Tan, Y. He, D. Yan, H.X. Luo, The effect of Fe/Co ratio on the structure and oxygen permeability of Ca-containing composite membranes, Inorg. Chem. Front. 6 (2019) 2885-2893. https://doi.org/10.1039/C9QI00822E.

[29] J.Y. Wang, Q.K. Jiang, D.D. Liu, L.M. Zhang, L.L. Cai, Y. Zhu, Z.W. Cao, W.P. Li, X.F. Zhu, W.S. Yang, Effect of inner strain on the performance of dual-phase oxygen permeable membranes, Journal of Membrane Science 644 (2022) 120142. https://doi.org/10.1016/j.memsci.2021.120142.

[30] M. Chivite Lacaba, A. Alveal, J. Prado-Gonjal, J.A. Alonso, M.T. Fernández Díaz, L. Troncoso, V. Cascos, Reducing the Cobalt Content in $SrCo_{0.95}Ti_{0.05}O_{3-\delta}$-Based Perovskites to Produce Cleaner Cathodes for IT-SOFCs, ACS Applied Energy Materials 6 (2023) 1046-1055. https://doi.org/10.1021/acsaem.2c03569.

[31] H. Luo, H. Jiang, K. Efimov, J. Caro, H. Wang, Influence of the preparation methods on the microstructure and oxygen permeability of a $CO_2$-stable dual phase membrane, AIChE Journal 57 (2011) 2738-2745. https://doi.org/10.1002/aic.12488.

[32] V. Kharton, A. Kovalevsky, E. Tsipis, A. Viskup, E. Naumovich, J. Jurado, J. Frade, Mixed conductivity and stability of A-site-deficient $Sr(Fe,Ti)O_{3-\delta}$ perovskites, Journal of Solid State Electrochemistry 7 (2002) 30-36. https://doi.org/10.1007/s10008-002-0286-3.

[33] Z. Shao, G. Xiong, J. Tong, H. Dong, W. Yang, Ba effect in doped $Sr(Co0.8Fe0.2)O3-\delta$ on the phase structure and oxygen permeation properties of the dense ceramic membranes, Separation and Purification Technology 25

(2001) 419-429. https://doi.org/10.1016/S1383-5866(01)00071-5.

[34] L. Shi, S. Wang, T.N. Lu, Y. He, D. Yan, Q. Lan, Z.A. Xie, H.Q. Wang, M.-R. Li, J. Caro, H.X. Luo, High $CO_2$-tolerance oxygen permeation dual-phase membranes $Ce_{0.9}Pr_{0.1}O_{2-\delta}$-$Pr_{0.6}Sr_{0.4}Fe_{0.8}Al_{0.2}O_{3-\delta}$, Journal of Alloys and Compounds 806 (2019) 500-509. https://doi.org/10.1016/j.jallcom.2019.07.281.

[35] Y.B. Liu, H.W. Cheng, Q.C. Sun, X.F. Xu, S. Chen, Q. Xu, X.G. Lu, Phase transition and oxygen permeability of $Pr_{0.6}Sr_{0.4}FeO_{3-\delta}$ ceramic membrane at high temperature, Journal of the European Ceramic Society 41 (2021) 1975-1983. https://doi.org/10.1016/j.jeurceramsoc.2020.10.064.

[36] M. Yashima, Crystal structures, structural disorders and diffusion paths of ionic conductors from diffraction experiments, Solid State Ionics 179 (2008) 797-803. https://doi.org/10.1016/j.ssi.2007.12.099.

[37] L. Navarrete, M. Balaguer, V.B. Vert, J.M. Serra, Tailoring Electrocatalytic Properties of Solid Oxide Fuel Cell Composite Cathodes Based on $(La_{0.8}Sr_{0.2})_{0.95}MnO_{3+\delta}$ and Doped Cerias $Ce_{1-x}Ln_xO_{2-\delta}$ (Ln = Gd, La, Er, Pr, Tb and $x$ = 0.1–0.2), Fuel Cells 17 (2017) 100-107. https://doi.org/10.1002/fuce.201600133.

[38] M. Balaguer, C. Solís, J.M. Serra, Study of the Transport Properties of the Mixed Ionic Electronic Conductor $Ce_{1-x}Tb_xO_{2-\delta}$ + Co ($x$ = 0.1, 0.2) and Evaluation As Oxygen-Transport Membrane, Chemistry of Materials 23 (2011) 2333-2343. https://doi.org/10.1021/cm103581w.

[39] D.P. Fagg, A.L. Shaula, V.V. Kharton, J.R. Frade, High oxygen permeability in fluorite-type $Ce_{0.8}Pr_{0.2}O_{2-\delta}$ via the use of sintering aids, Journal of Membrane Science 299 (2007) 1-7. https://doi.org/10.1016/j.memsci.2007.04.020.

[40] C. Chatzichristodoulou, P.V. Hendriksen, Electronic conductivity of $Ce_{0.9}Gd_{0.1}O_{1.95-\delta}$ and $Ce_{0.8}Pr_{0.2}O_{2-\delta}$: Hebb–Wagner polarisation in the case of redox active dopants and interference, Physical Chemistry Chemical Physics 13 (2011) 21558-21572. http://dx.doi.org/10.1039/C1CP21824G.

[41] C. Zhang, Y. Zhu, X.P. Wang, Y.H. Huang, L.Y. Zeng, K. Li, P.F. Yu, K.W. Wang, L.F. Li, Z.C. Xiang, R. Chen, X.F. Zhu, H.X. Luo, Excellent and $CO_2$-resistant permeability of $Ce_{0.85}Nd_{0.1}Cu_{0.05}O_{2-\delta}$-$Nd_xSr_{1-x}Fe_{1-y}Cu_yO_{3-\delta}$ dual-phase oxygen transport membranes, Journal of Membrane Science 696 (2024) 122485. https://doi.org/10.1016/j.memsci.2024.122485.

[42] X.P. Wang, Y.H. Huang, D.C. Li, L.Y. Zeng, Y.Y. He, M. Boubeche, H.X. Luo, High oxygen permeation flux of cobalt-free Cu-based ceramic dual-phase membranes, Journal of Membrane Science 633 (2021) 119403. https://doi.org/10.1016/j.memsci.2021.119403.

[43] G.X. Chen, W.M. Liu, M. Widenmeyer, P.J. Ying, M.F. Dou, W.J. Xie, C. Bubeck, L. Wang, M. Fyta, A. Feldhoff, A. Weidenkaff, High flux and $CO_2$-resistance of $La_{0.6}Ca_{0.4}Co_{1-x}Fe_xO_{3-\delta}$ oxygen-transporting membranes, Journal of Membrane Science 590 (2019) 117082. https://doi.org/10.1016/j.memsci.2019.05.007.

[44] J.W. Zhu, S.B. Guo, Z.Y. Chu, W.Q. Jin, $CO_2$-tolerant oxygen-permeable perovskite-type membranes with high permeability, J. Mater. Chem. A 3 (2015)

22564-22573. https://doi.org/10.1039/C5TA04598C.

[45] W. Fang, F. Steinbach, C.S. Chen, A. Feldhoff, An Approach To Enhance the $CO_2$ Tolerance of Fluorite–Perovskite Dual-Phase Oxygen-Transporting Membrane, Chemistry of Materials 27 (2015) 7820-7826. https://doi.org/10.1021/acs.chemmater.5b03823.

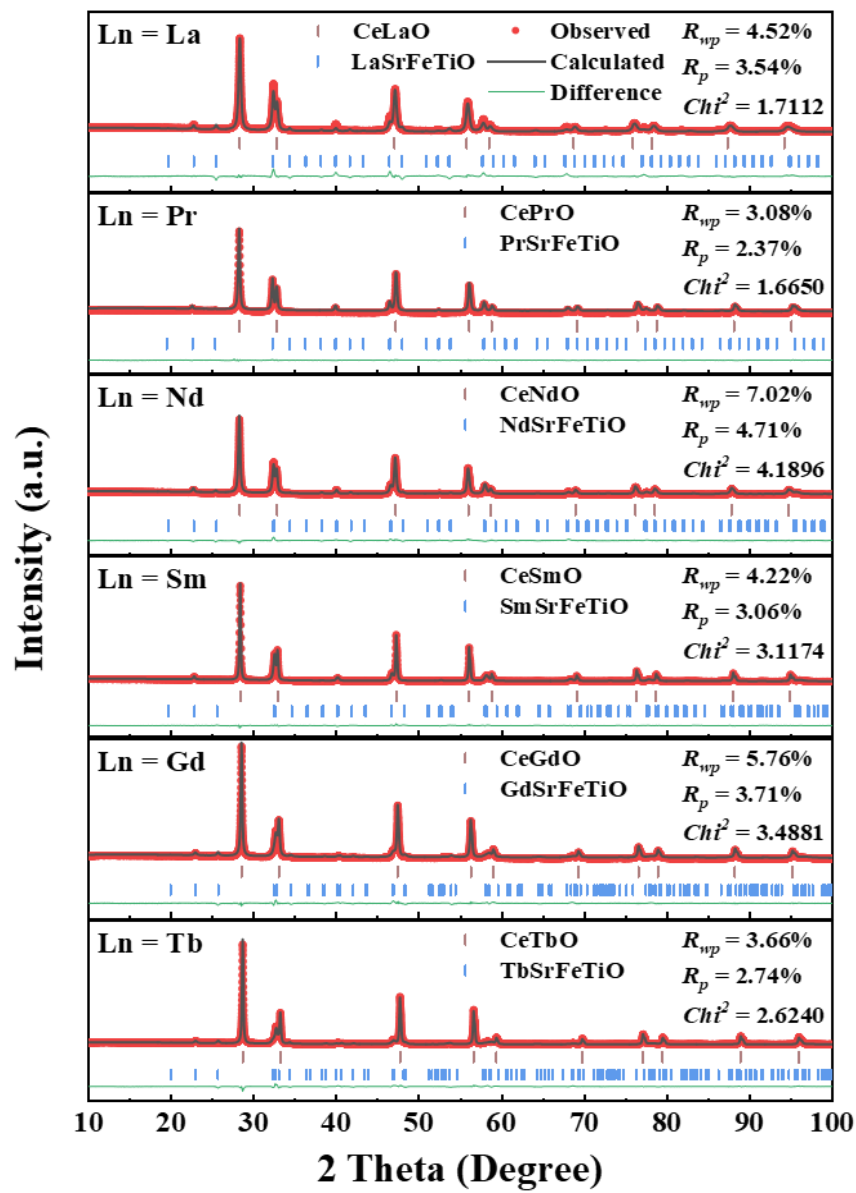

**Fig. 1.** XRD refinements of CLnO-LnSFTO (Ln = La, Pr, Nd, Sm, Gd, Tb) powders after 950 ºC calcination.

**Table 1.** Cell parameters of CLnO-LnSFTO (Ln = La, Pr, Nd, Sm, Gd, Tb) powders calcined at 950 °C obtained by XRD data refinement with Rietveld model.

| Ln | Fluorite (*No.* 225: Fm-3m) | Perovskite (*No.* 62: Pbnm) | | | |
|---|---|---|---|---|---|
|  | $a$ / Å | $a$ / Å | $b$ / Å | $c$ / Å | $\Phi$ / ° |
| **La** | 5.5314 (8) | 5.6103 (9) | 5.6154 (8) | 7.9432 (12) | 3.66 |
| **Pr** | 5.3989 (19) | 5.4562 (8) | 5.4741 (9) | 7.7177 (13) | 4.77 |
| **Nd** | 5.4613 (14) | 5.5398 (9) | 5.5549 (7) | 7.8495 (8) | 5.52 |
| **Sm** | 5.4276 (5) | 5.4753 (9) | 5.5050 (8) | 7.7582 (13) | 6.93 |
| **Gd** | 5.4221 (4) | 5.4562 (11) | 5.5218 (9) | 7.7477 (15) | 10.23 |
| **Tb** | 5.3879 (8) | 5.4256 (6) | 5.5454 (8) | 7.7361 (9) | 13.97 |

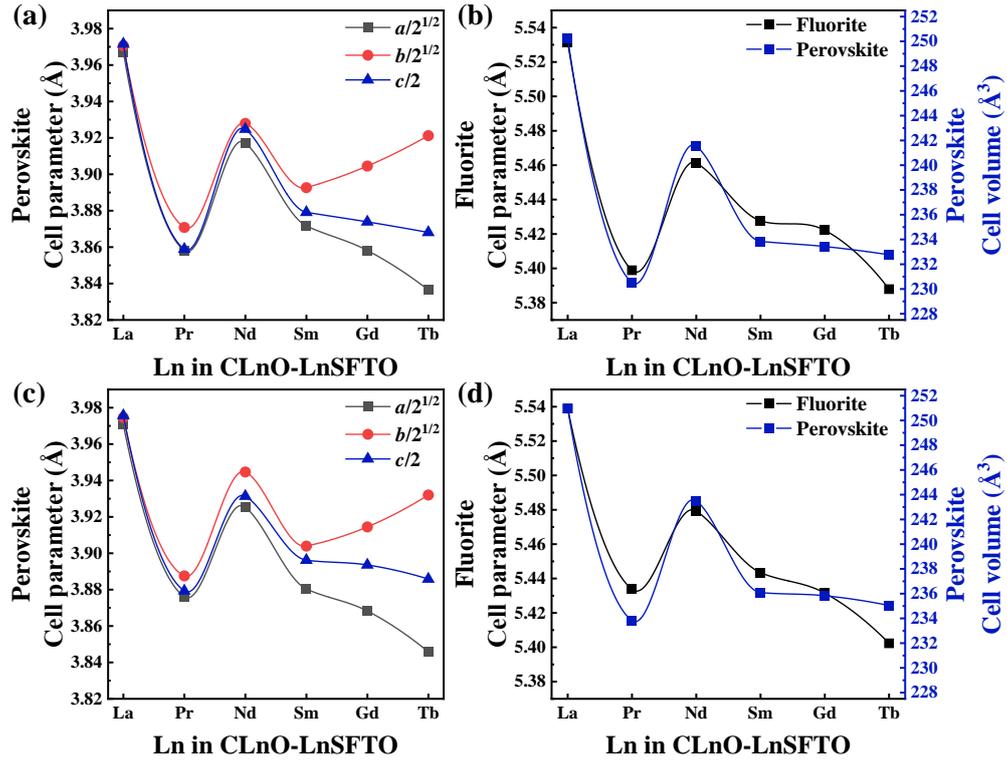

**Fig. 2.** The evolution of the cell parameters for CLnO-LnSFTO (a, b) powders and (c, d) membranes.

**Table 2.** Cell parameters of CLnO-LnSFTO (Ln = La, Pr, Nd, Sm, Gd, Tb) membranes sintered at 1450 ºC for 5 hours, obtained by XRD data refinement with the Rietveld model.

| Ln | Fluorite (*No.* 225：Fm-3m) | Perovskite (*No.* 62：Pbnm) | | | |
|---|---|---|---|---|---|
| | *a* / Å | *a* / Å | *b* / Å | *c* / Å | *Φ* / ° |
| La | 5.5389 (6) | 5.6153 (7) | 5.6209 (8) | 7.9513 (11) | 3.85 |
| Pr | 5.4337 (7) | 5.4816 (8) | 5.4978 (6) | 7.7586 (15) | 4.98 |
| Nd | 5.4793 (9) | 5.5513 (7) | 5.5786 (8) | 7.8627 (12) | 6.49 |
| Sm | 5.4434 (5) | 5.4876 (9) | 5.5210 (8) | 7.7923 (14) | 8.15 |
| Gd | 5.4315 (8) | 5.4705 (6) | 5.5359 (7) | 7.7871 (10) | 10.96 |
| Tb | 5.4023 (10) | 5.4387 (11) | 5.5607 (10) | 7.7716 (19) | 14.54 |

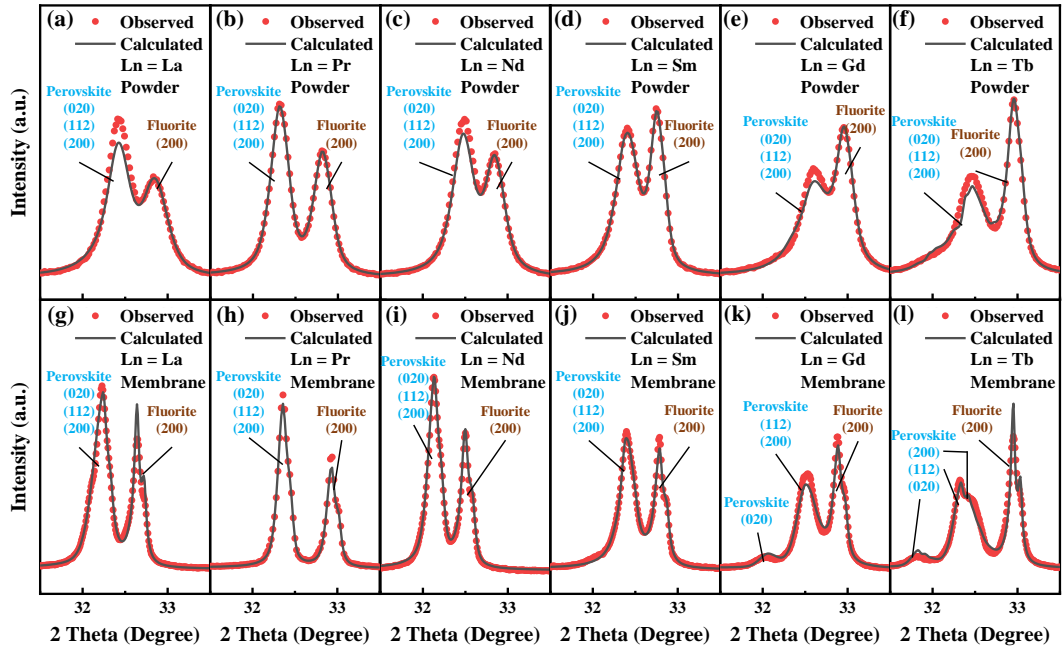

**Fig. 3.** The close-up of XRD patterns of CLnO-LnSFTO (Ln = La, Pr, Nd, Sm, Gd, Tb) powder and membranes near 32°.

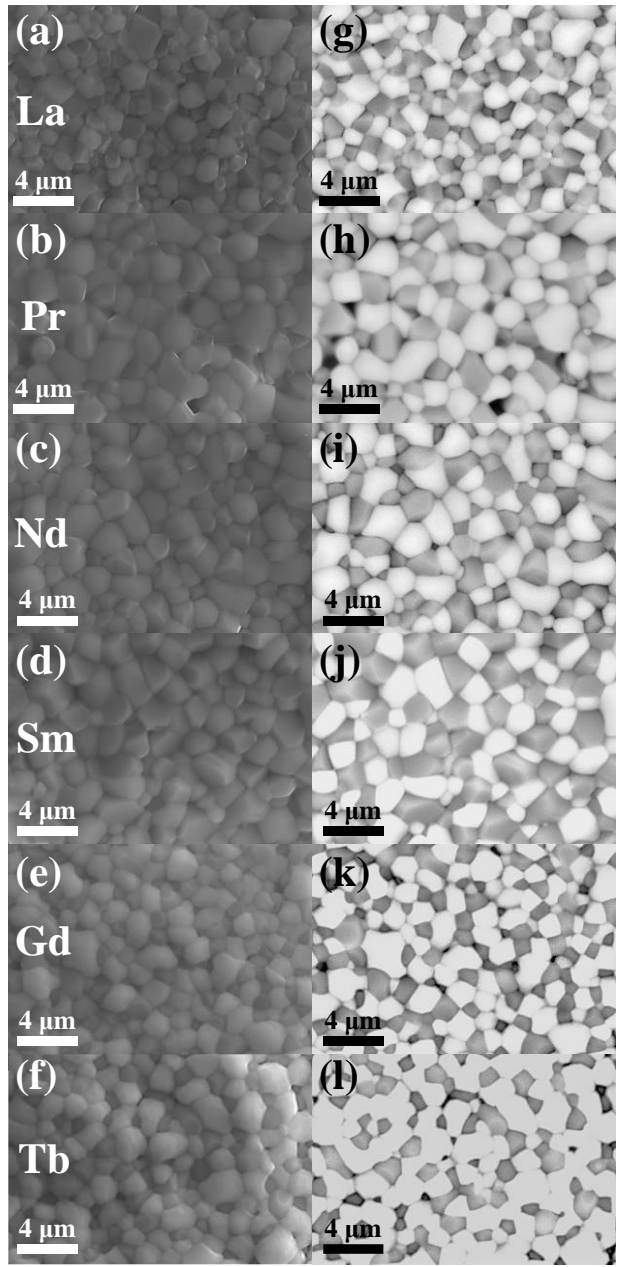

**Fig. 4.** (a-f) SEM and (g-l) BSEM images of surface morphology of CLnO-LnSFTO (Ln = La, Pr, Nd, Sm, Gd, Tb) fresh membranes sintered at 1450 ºC.

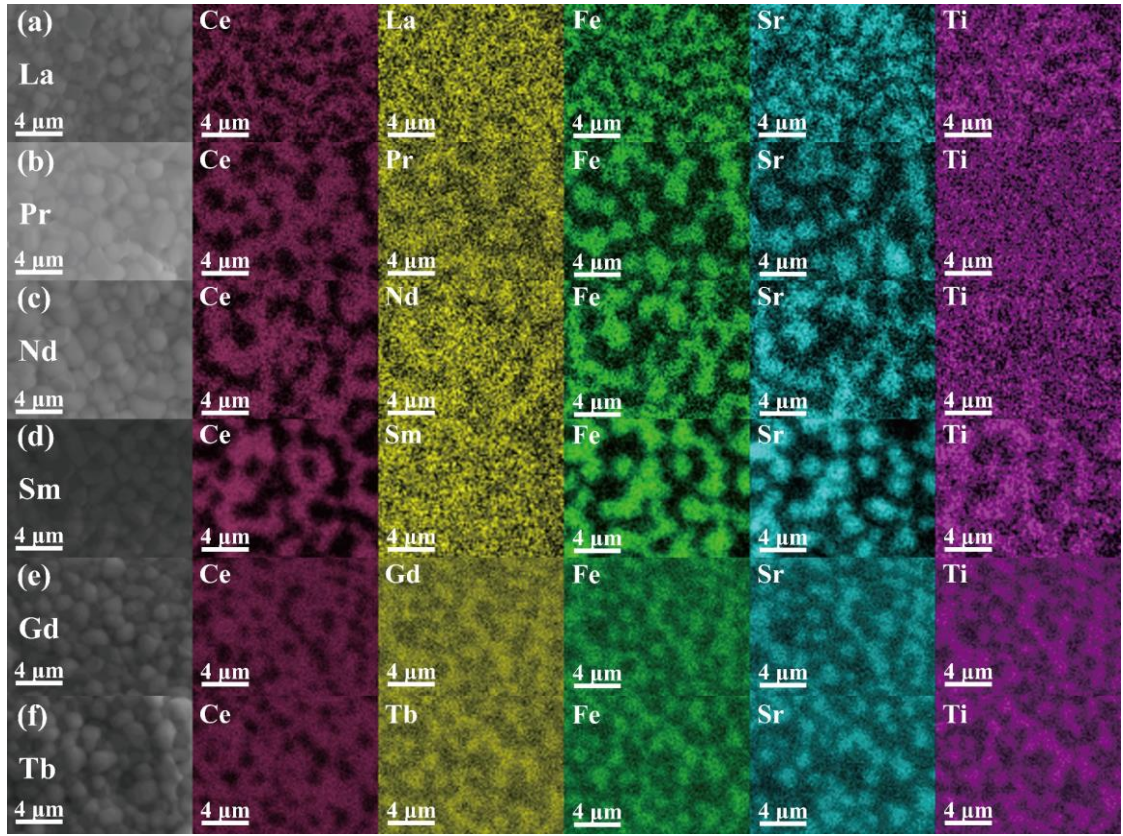

**Fig. 5.** EDXS images of elements on the surface of CLnO-LnSFTO (Ln = La, Pr, Nd, Sm, Gd, Tb) fresh membranes sintered at 1450 °C.

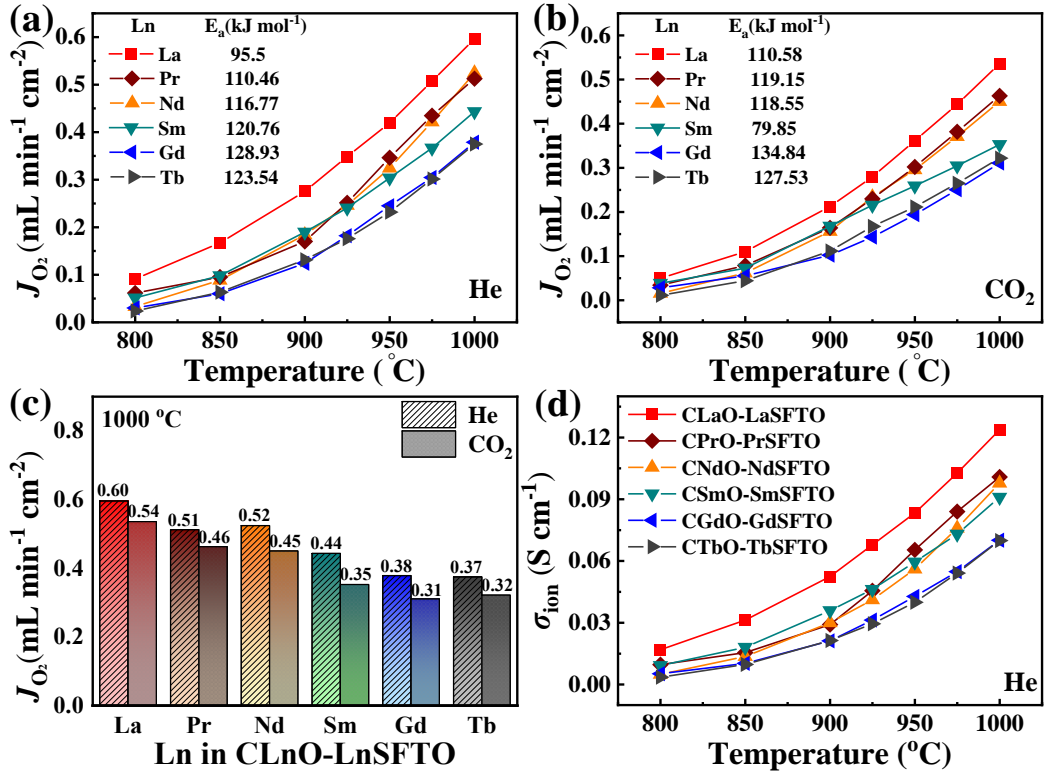

**Fig. 6.** Curves of oxygen permeation fluxes of CLnO-LnSFTO (Ln = La, Pr, Nd, Sm, Gd, Tb) membranes versus temperature with (a) He and (b) $CO_2$ sweeping. (c) Comparison of oxygen permeation fluxes. (d) Ionic conductivity of CLnO-LnSFTO (Ln = La, Pr, Nd, Sm, Gd, Tb) membranes calculated with **Eq. (5)**.

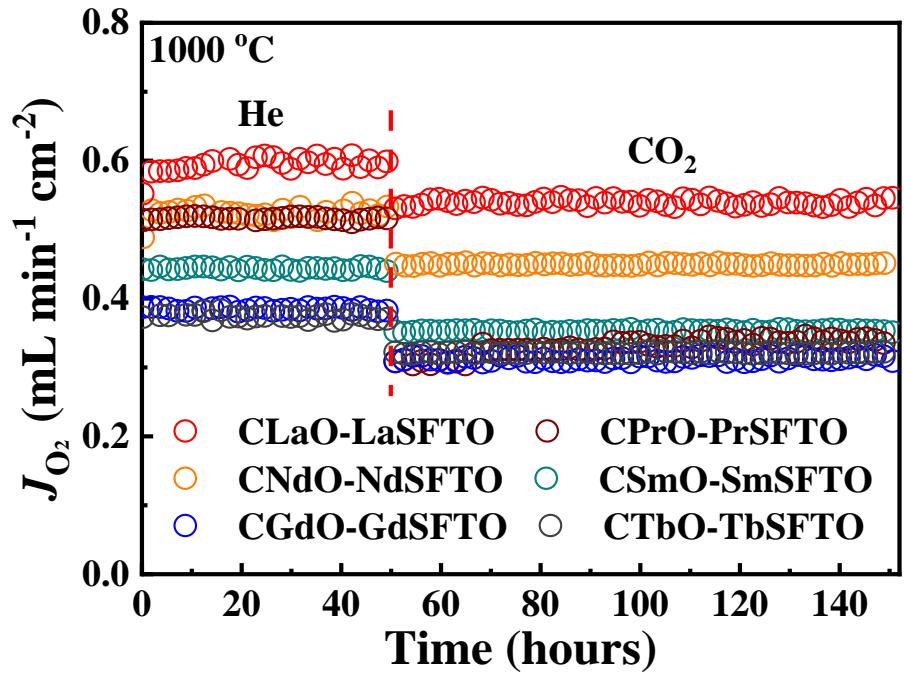

**Fig. 7.** Oxygen permeation fluxes through CLnO-LnSFTO (Ln = La, Pr, Nd, Sm, Gd, Tb) membranes during long-term tests at 1000 °C.

# Supplementary material

# Effects of Lanthanides on the Structure and Oxygen Permeability of Ti-doped Dual-phase Membranes

*Chao Zhang, Zaichen Xiang, Lingyong Zeng, Peifeng Yu, Kuan Li, Kangwang Wang, Longfu Li, Rui Chen, Huixia Luo\**

School of Materials Science and Engineering, State Key Laboratory of Optoelectronic Materials and Technologies, Guangdong Provincial Key Laboratory of Magnetoelectric Physics and Devices, Key Lab of Polymer Composite & Functional Materials, Sun Yat-Sen University, No. 135, Xingang Xi Road, Guangzhou, 510275, P. R. China

*[\*]Corresponding author/authors complete details (Telephone; E-mail:) (+0086)-2039386124*

*luohx7@mail.sysu.edu.cn*

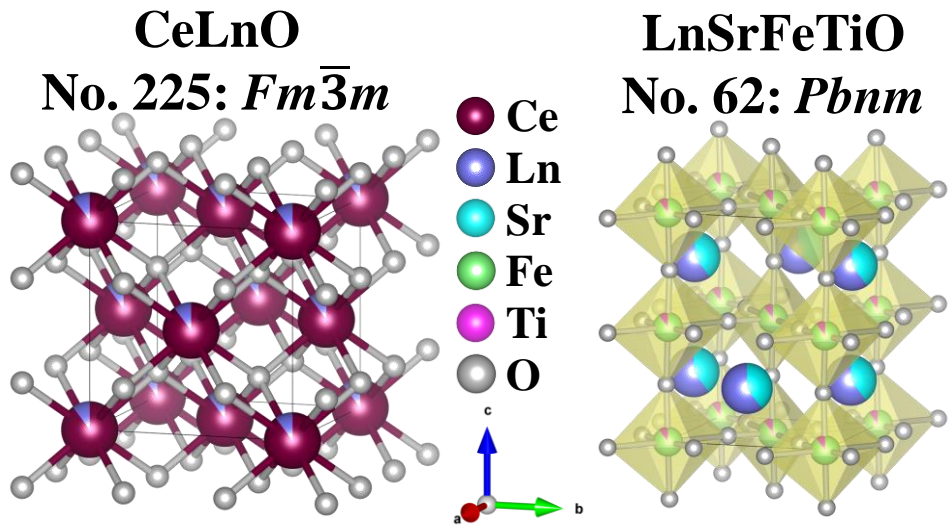

Fig. S1. Graphs of the crystal structure of the CLnO phase and the LnSFTO phase (Ln = La, Pr, Nd, Sm, Gd, Tb).

**Table S1.** Ionic radii of Ln (Ln = La, Pr, Nd, Sm, Gd, Tb) elements with different coordination numbers [27].

| Phase (Coordination number) | Cation | Ionic radii (pm) |
| --- | --- | --- |
| Fluorite (8) | $Ce^{4+}$ | 97 |
| | $Ce^{3+}$ | 114.3 |
| | $La^{3+}$ | 116 |
| | $Pr^{4+}$ | 96 |
| | $Pr^{3+}$ | 112.6 |
| | $Nd^{3+}$ | 110.9 |
| | $Sm^{3+}$ | 107.9 |
| | $Gd^{3+}$ | 106 |
| | $Tb^{4+}$ | 88 |
| | $Tb^{3+}$ | 104 |
| Perovskite (12) | $La^{3+}$ | 136 |
| | $Pr^{3+}$ | 132 |
| | $Nd^{3+}$ | 127 |
| | $Sm^{3+}$ | 124 |
| | $Gd^{3+}$ | 127 |
| | $Tb^{3+}$ | 125 |

**Reference**

[27] Y.Q. Jia, Crystal radii and effective ionic radii of the rare earth ions, J. Solid State Chem. 95 (1991) 184-187. https://doi.org/10.1016/0022-4596(91)90388-X.

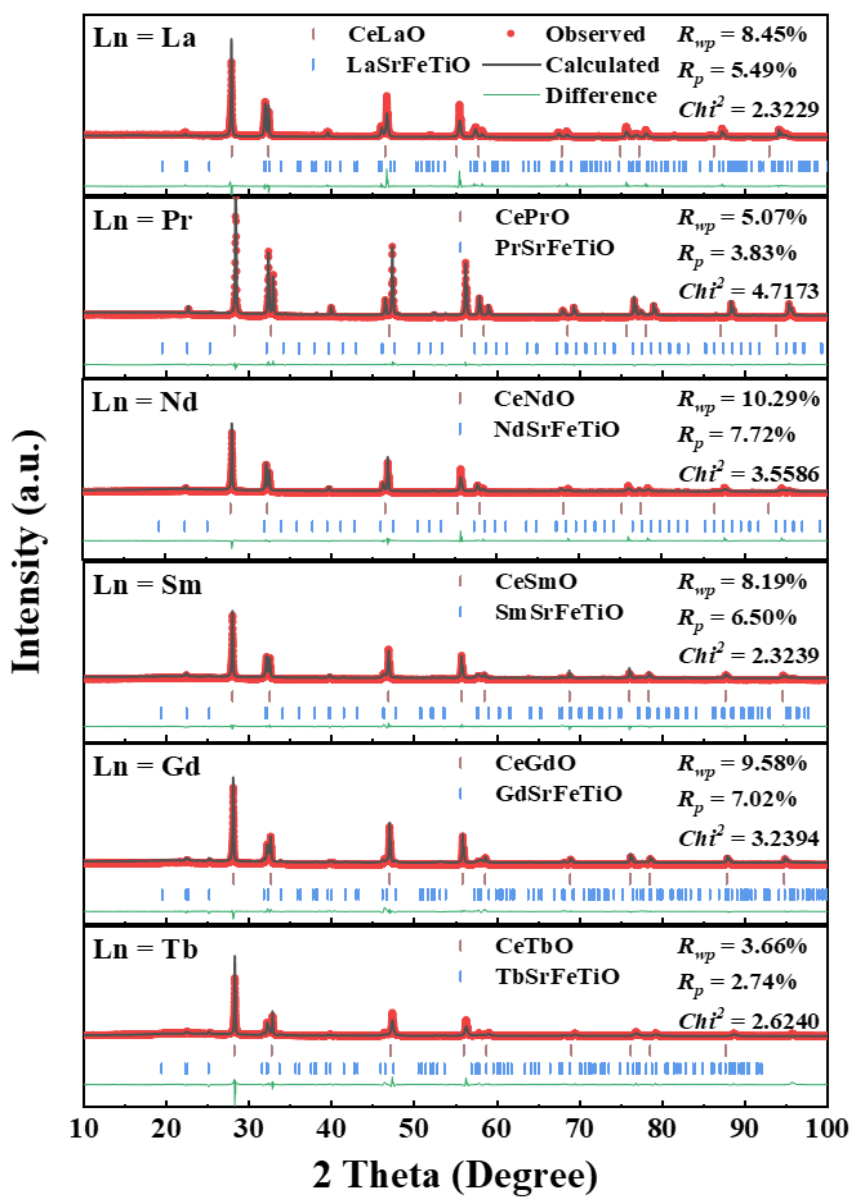

**Fig. S2.** XRD refinements of the CLnO-LnSFTO (Ln = La, Pr, Nd, Sm, Gd, Tb) membranes sintered at 1450 °C for 5 hours.

**Table S2.** The experimental and theoretical atomic composition on the surface of fresh CLnO-LnSFTO (Ln = La, Pr, Nd, Sm, Gd, Tb) membranes, characterized by EDXS.

| Ln | Type | Element (Atomic percentage) | | | | |
| --- | --- | --- | --- | --- | --- | --- |
| | | Ce | Fe | Ln | Ti | Sr |
| La | Theoretical | 44.213 | 22.894 | 20.175 | 2.544 | 10.175 |
| | Experimental | 51.227 | 20.441 | 14.677 | 2.166 | 11.485 |
| Pr | Theoretical | 44.227 | 22.886 | 20.172 | 2.543 | 10.172 |
| | Experimental | 42.715 | 22.357 | 19.737 | 2.096 | 13.095 |
| Nd | Theoretical | 44.466 | 22.767 | 20.119 | 2.530 | 10.119 |
| | Experimental | 49.685 | 19.981 | 18.257 | 1.691 | 10.387 |
| Sm | Theoretical | 44.751 | 22.625 | 20.055 | 2.514 | 10.055 |
| | Experimental | 47.895 | 22.646 | 17.229 | 2.580 | 9.650 |
| Gd | Theoretical | 45.065 | 22.467 | 19.986 | 2.496 | 9.986 |
| | Experimental | 53.632 | 19.211 | 17.013 | 2.134 | 8.010 |
| Tb | Theoretical | 45.141 | 22.430 | 19.969 | 2.492 | 9.969 |
| | Experimental | 55.790 | 18.334 | 17.674 | 1.575 | 6.628 |

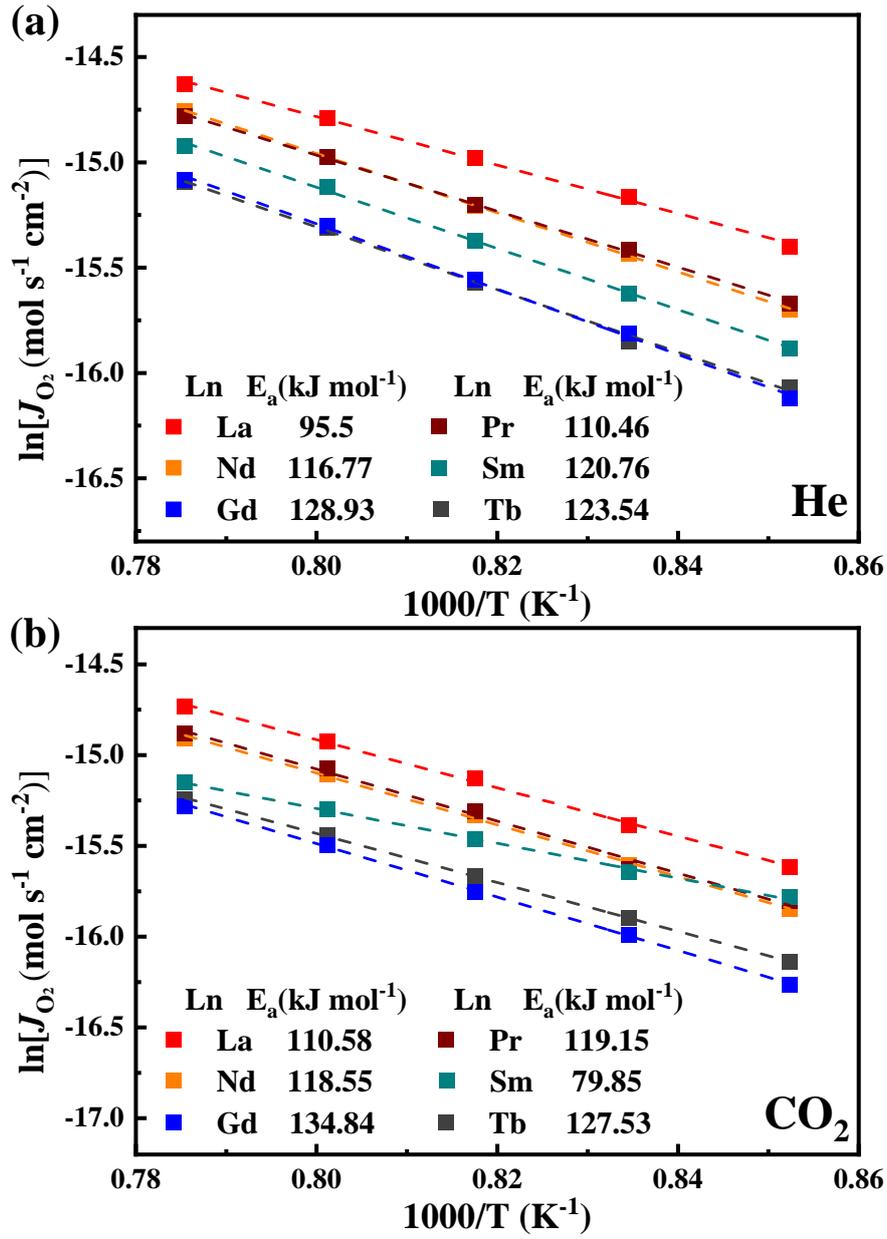

**Fig. S3.** Arrhenius plots of oxygen permeation fluxes and activation energy fitting of CLnO-LnSFTO (Ln = La, Pr, Nd, Sm, Gd, Tb) OTMs with (a) He and (b) $CO_2$ sweeping.

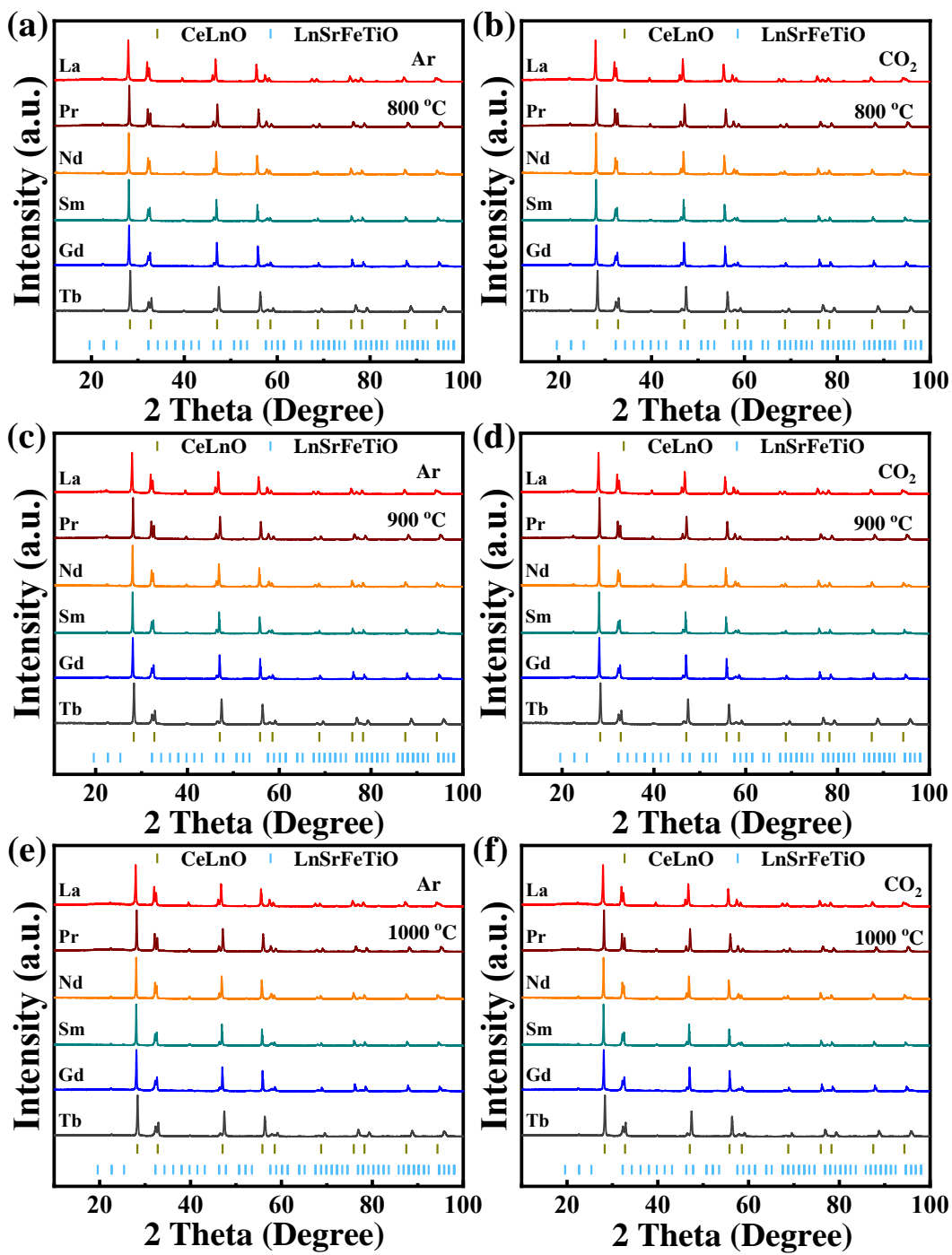

**Fig. S4.** XRD patterns of CLnO-LnSFTO (Ln = La, Pr, Nd, Sm, Gd, Tb) powders treated in (a, c, e) pure Ar and (b, d, f) pure $CO_2$ at (a, b) 800 ºC, (c, d) 900 ºC, and (e, f) 1000 ºC for 24 hours.

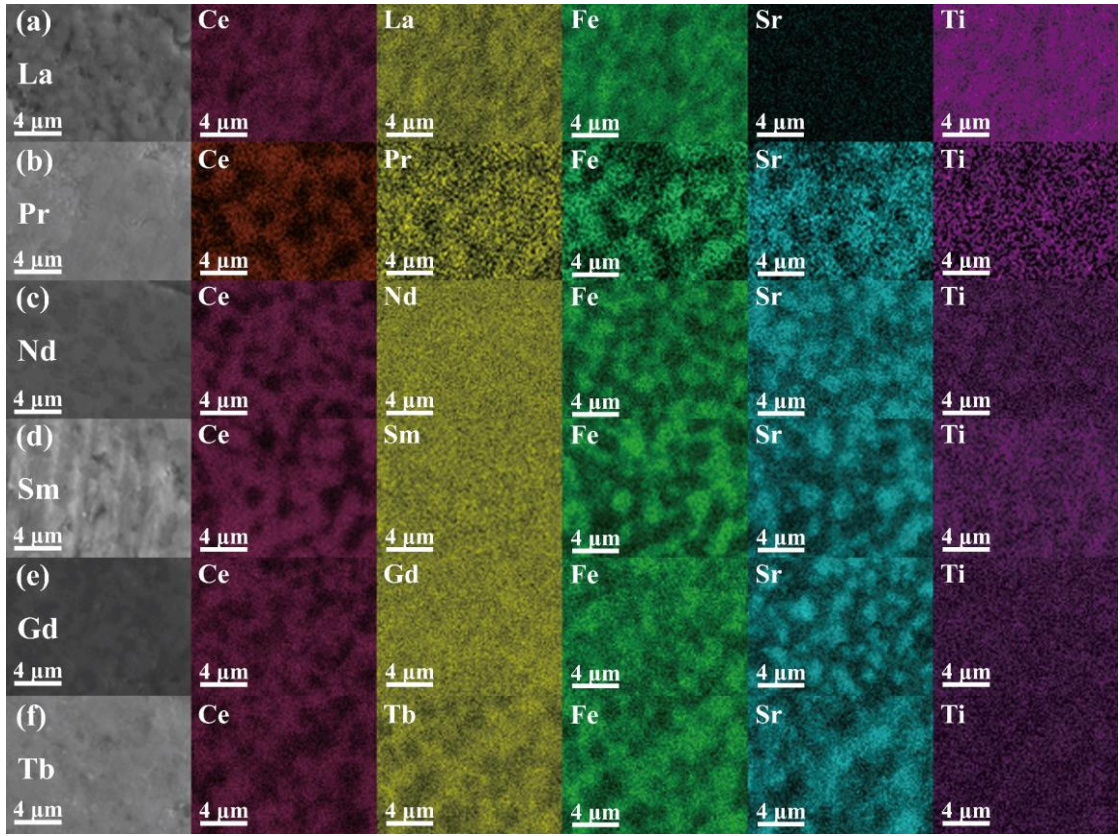

**Fig. S5.** SEM-EDXS images on the feed side of the spent CLnO-LnSFTO (Ln = La, Pr, Nd, Sm, Gd, Tb) membranes after long-term permeation tests.

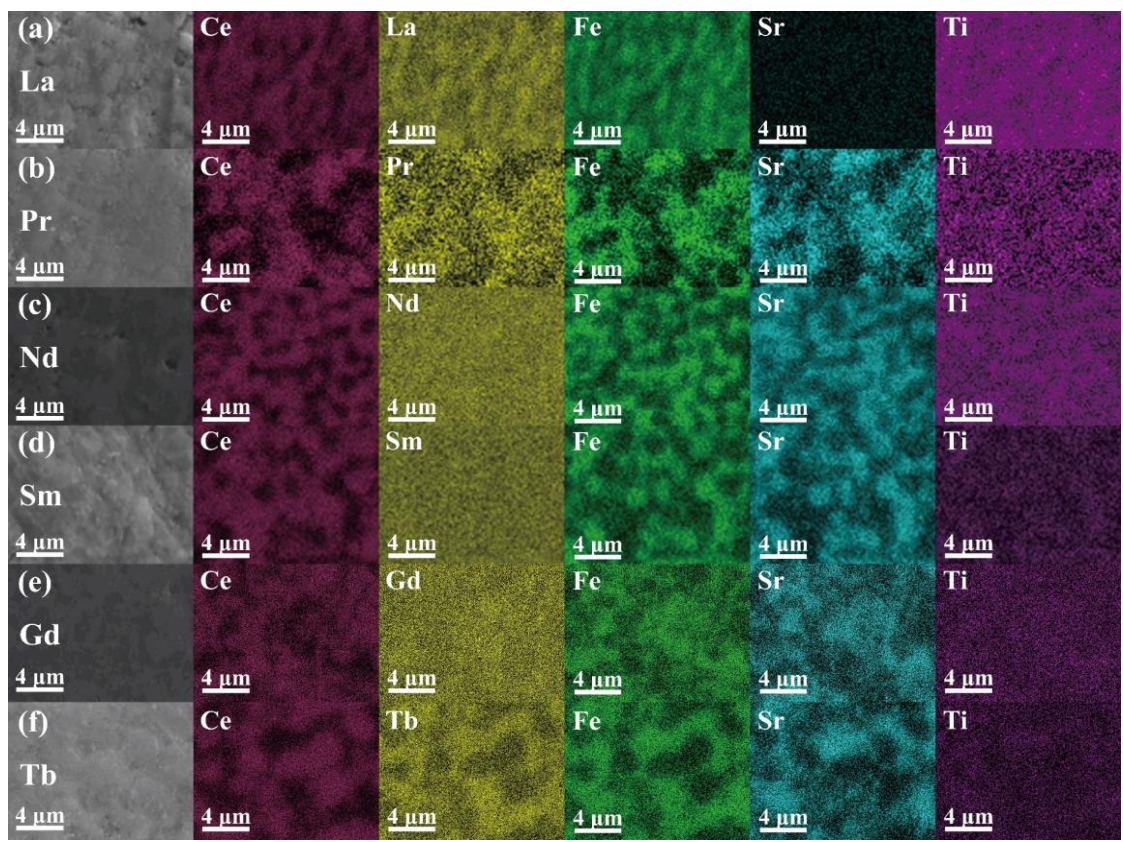

**Fig. S6.** SEM-EDXS images on the feed side of spent CLnO-LnSFTO (Ln = La, Pr, Nd, Sm, Gd, Tb) membranes after long-term permeation tests.

**Table S3.** The atomic percentage of elements on the feed side and sweep side of tested CLnO-LnSFTO (Ln = La, Pr, Nd, Sm, Gd, Tb) membranes, characterized by EDXS.

| Ln | Side | Element (Atomic percentage) | | | | |
|---|---|---|---|---|---|---|
| | | **Ce** | **Fe** | **Ln** | **Ti** | **Sr** |
| **La** | Feed | 52.405 | 19.883 | 13.304 | 3.715 | 10.693 |
| | Sweep | 51.527 | 20.411 | 13.396 | 3.636 | 11.031 |
| **Pr** | Feed | 40.978 | 24.323 | 20.175 | 1.850 | 12.673 |
| | Sweep | 41.687 | 23.889 | 20.022 | 2.055 | 12.347 |
| **Nd** | Feed | 44.144 | 22.378 | 18.329 | 3.109 | 12.041 |
| | Sweep | 43.523 | 22.737 | 18.580 | 3.091 | 12.068 |
| **Sm** | Feed | 46.499 | 21.458 | 17.842 | 2.261 | 11.940 |
| | Sweep | 44.343 | 22.181 | 17.729 | 2.470 | 12.687 |
| **Gd** | Feed | 41.293 | 18.934 | 17.874 | 2.226 | 19.674 |
| | Sweep | 43.098 | 22.692 | 19.112 | 2.518 | 12.580 |
| **Tb** | Feed | 43.170 | 18.581 | 22.982 | 1.781 | 13.486 |
| | Sweep | 39.092 | 21.115 | 22.635 | 2.068 | 15.090 |